\documentclass[epsfig]{article}
\usepackage{graphics}
\begin{document}
\newcommand{\beq}{\begin{equation}}
\newcommand{\eeq}{\end{equation}}
\newcommand{\beqn}{\begin{eqnarray}}
\newcommand{\eeqn}{\end{eqnarray}}
\newcommand{\dpf}{\displaystyle\frac}
\newcommand{\no}{\nonumber}
\newcommand{\ep}{\epsilon}
\begin{center}
{\Large Holography in (2+1)-dimensional cosmological models}
\end{center}
\vspace{1ex}
\centerline{\large Bin
Wang$^{a,b,}$\footnote[1]{e-mail:binwang@fma.if.usp.br},
\ Elcio Abdalla$^{a,}$\footnote[2]{e-mail:eabdalla@fma.if.usp.br}}
\begin{center}
{$^{a}$ Instituto De Fisica, Universidade De Sao Paulo, C.P.66.318, CEP
05315-970, Sao Paulo, Brazil \\
$^{b}$ Department of Physics, Shanghai Teachers' University, P. R. China}
\end{center}
\vspace{6ex}
\begin{abstract}
The cosmic holographic principle suggested by Fischler and
Susskind has been examined in (2+1)-dimensional cosmological models. 
Analogously to the (3+1)-dimensional counterpart, the holographic
principle is satisfied in all flat and open universes. For (2+1)-dimensional
closed universes the holographic principle cannot be realized in general.
The principle  cannot be maintained, neither  introducing negative 
pressure matter nor matter with highly unconventional equation of state.
\end{abstract}
\vspace{6ex}
\hspace*{0mm} PACS number(s): 04.70.Dy, 98.80.Cq
\vfill
\newpage
It is well-known that the entropy of a black hole is represented in
terms of the area $A$ of the
event horizon by the formula $S=A/4$. This relation can be interpreted
as meaning that all the degrees of freedom
in a black hole are stored on its boundary. 
For a long time one has sought to extend this relation to more general
situations in gravity,
meaning that under certain conditions all the information contained in a
physical
system in quantum gravity can be projected onto its boundary. This is
called the holographic
principle \cite{1,2}. Recently this hope has been further promoted by the
discovery
of AdS/CFT (Anti-de Sitter/Conformal Field Theory) correspondence \cite{3}.
The conjectured equivalence between
Supergravity on D-dimensional AdS space and conformal field theory on
D-1-dimensional boundary has been proved very useful to
gain information about
supergravity in the bulk from knowledge about its boundary. It is
generally believed that
the holographic principle must eventually have implications for cosmology.

There have been many attempts to apply different formulations of the
holographic principle to various cosmological models \cite{4}--\cite{12}. 
A specific realization was suggested by Fischler and Susskind \cite{4}. 
A remarkable point
of their proposal is that the holographic principle is valid for  flat
or open universes with the equation of state satisfying the
condition $0\leq P\leq\rho$. However, for the closed universe the
principle is violated. The problem becomes even more serious if one
investigates the universe with a negative cosmological constant \cite{5}. In
that case the holographic principle fails, independently of whether the
universe is closed, open or flat. Various different modifications of
Fischler and Susskind's version of holographic principle have been raised 
recently, such as replacing the holographic principle by the generalized
second law of thermodynamics \cite{6,5}, using cosmological apparent horizon
instead of particle horizon in the formulation of holographic principle
\cite{7}, changing the definition of ``degrees of freedom" 
\cite{8} etc. A very recent result claimed that the holographic principle 
in a closed universe can be obeyed if the universe contains strange 
negative pressure matter \cite{12}. All these
studies have concentrated on (3+1)-dimensional cosmology and it
would be fair to say that the fully understanding of the cosmic
holography is still lacking.

The purpose of the present paper is to reconsider the previous studies in
(2+1)-dimensional cosmological models, where the mathematical
simplicity  may help us 
understanding the physics deeper. The fact that the first successful
implication of AdS/CFT duality to solve the problem of the microscopic
interpretation of black hole entropy appeared in (2+1)-dimensional models
[13-16], giving more hints in that of higher-dimensional black hole
cases
[17-19], gives us further motivation to study (2+1)-dimensional
cosmological models. We will try to examine the basic assumptions of
cosmic holography and show that for open and flat universes with a certain 
fixed
equation of state, the result in (2+1)-dimensional cases reproduces
that of (3+1)-dimensional cosmology [4-6]. Applying the method proposed 
in \cite{5}, we find that the holographic principle in (2+1)-dimensional
closed universes containing normal matter of positive pressure cannot be
realized. Attempts to protect the holographic principle by introducing
negative pressure matter as well as matter with highly unconventional
state equation also has been foiled.

We first consider the equations governing homogeneous expansion in
(2+1)-dimensions [20]. In terms of the (2+1)-dimensional Robertson-Walker
line
element
\beq     
{\rm d}s^2={\rm d}t^2-a^2(t)(\dpf{{\rm d}r^2}{1-kr^2}+r^2{\rm d}\theta^2),
\eeq
Einstein field equations become
\beqn  
(\dpf{\dot{a}}{a})^2+\dpf{k}{a^2} & = & 2\pi G\rho, \\
\dpf{\ddot{a}}{a} & = & -2\pi GP, \\
\dpf{d}{dt}(\rho a^2)+P\dpf{d}{dt}a^2 & = & 0.
\eeqn
For a dust-filled universe, $P=0$ and energy-momentum conservation leads
to
\beqn 
\rho a^2 & = & const=\rho_0a^2 _0 \\
\dot{a}^2 & = & 2GM_0 -k
\eeqn
where $M_0 =\pi\rho_0 a^2 _0$. Eq(6) requires $M_0 \geq k/2G$, therefore
the solution 
\beq    
a(t)=a_0+\sqrt{2GM_0 -k}t
\eeq
implies that the universe will always expand, regardless of the value
of $k$.

The comoving distance to the horizon $r_H$ is defined by
\beq 
r_H=\int^t_0\dpf{dt'}{a(t')}=\dpf{1}{\sqrt{2GM_0 -k}}
[\ln(a_0+\sqrt{2GM_0 -k}t)-\ln a_0],
\eeq
while the corresponding physical distance is $L_H=a(t)r_H$. The total
entropy  inside the horizon divided by the area grows as 
\beq  
\dpf{S}{A}=\sigma\dpf{r_H}{a}=\dpf{\sigma}{\sqrt{2GM_0 
-k}}\dpf{1}{a}\ln\dpf{a}{a_0},
\eeq
where $\sigma$ is the comoving entropy density and stisfies $\sigma \sim
a_0^2$ at the beginning.
Thus, later on, as the universe expands, the bound  will be
satisfied even
better. This result is in agreement with the case of the flat and open
universes discussed in (3+1)-dimensional cases [4-6].

Let us consider now a radiation-dominated universe [20], namely 
$\rho=2P$. Energy-momentum conservation now implies that
\beqn     
\rho a^3 & = & const=\rho_0a_0^3, \\
\dot{a}^2 & = & 2GM_0/a - k, \\
\ddot{a} & = & - M_0 G/a^2.
\eeqn
In the flat case $(k=0)$, the solution of (11) is straightforward
\beq  
a(t)=[\dpf{3}{2}(2GM_0)^{1/2}]^{2/3}t^{2/3}.
\eeq
The comoving horizon is 
\beq     
r_H=\int^t_0\dpf{dt'}{a(t')}=\dpf{3}{[\dpf{3}{2}(2GM_0)^{1/2}]^{2/3}}
t^{1/3}.
\eeq
For large $t$, the comoving entropy area relation behaves as 
\beq  
\dpf{S}{A}\sim \dpf{2\sigma}{(2GM_0)^{1/2}a^{1/2} (t)}.
\eeq
Thus the ratio $S/A$ does not increase with time. Therefore if the initial
holographic constraint was satisfied at the beginning, namely,
$\sigma/a^2_0\leq 1$, as the universe expands the holographic bound will
get more stringent, which agrees with the (3+1)-dimensional results [4-6].

The behavior of isotropic open universe $(k=-1)$ is similar to that of the
flat universe.

It is of interest to discuss the closed universe model $(k=1)$ in the
radiation-dominated phase. First of all, we see from Eq(11) that
$\dot{a}$ vanishes at $a=2GM_0 $, after which $\dot{a}$ becomes
negative
and universe collapses. This happens within a finite time after the
beginning of the expansion. From the definition of the particle horizon
and (11), one can find the value of $L_H$ at the turning point,
\beq 
L_H(turning)=2GM_0 B(1/2,1/2),
\eeq
where $B(p,q)$ is the Euler beta function. Putting these formulas
together,
we see that at the turning point 
\beq  
\dpf{S}{A}\sim \dpf{1}{4GM_0}. \label{17}
\eeq
The ratio $S/A$ is
seen to be smaller than unit at the turning point. Now we can consider
what happens
near the final stages of collapse, when the universe has evolved back to 
the Planckian
scale. By symmetry, $L_H\sim \dpf{2a_0}{a(turning)}L_H(turning)\sim
2a_0$ [5]. The scale factor
for the closed universe has been calculated in [20]
\beq 
t=2GM_0[\arcsin(\dpf{a}{2GM_0})^{1/2}-(\dpf{a}{2GM_0})^{1/2}
(1-\dpf{a}{2GM_0})^{1/2}]. \no
\eeq
Considering the small value of $a$ at the Planck time $t=1$, we can expand
the above equation and we have
\beq     
a_0=[2(2GM_0)^{1/2}]^{2/3}.
\eeq
At this time, $\sigma/a^2_0 \sim 1$,
hence the ratio $S/A$ yields
\beq   
\dpf{S}{A}\sim 2a_0 \label{soversima0}
\eeq
We see that in order to have $S/A<1$, we need
$a_0<1/2$. However, from Eqs. (\ref{17}) and (\ref{soversima0}) we
learn that at the turning point
\begin{eqnarray}
\frac SA = \frac 2{a_0^3}\quad ,
\end{eqnarray}
thus the  holographic bound is violated at the turning point, which means
that the holographic bounds are violated much before the universe reaches
the future Planck era.

In order to generalize our discussion, we study in the
following the (2+1)-dimensional cosmological model proposed in [21]. The
metric is determined by a time-dependent scale factor $a(t)$ and of the
form Eq.(1). 
The scale
factor is determined by a Friedmann-like equation 
\beq 
(\dpf{\dot{a}}{a})^2=\dpf{2GM_0}{a^{2\gamma}}-\dpf{k}{a^2}\quad ,
\label{22-1}
\eeq 
when the material content is a perfect fluid with
equation of state $P=(\gamma -1)\rho$, where $\gamma$ is a constant. For
$1<\gamma\leq 2$, the solutions of (22) are closed, open or flat
cosmological models according to whether $k$ is 1, -1, or 0, respectively.
For $\gamma <1$ the universe is filled with negative pressure matter,
a case which was discussed in four-dimensional space-time, where it
has been argued that the holographic principle can be maintained \cite{12}.
The case $\gamma >2$, though not physical, is going to be discussed as well.

We will first discuss the flat cosmological model with $k=0$. From 
(\ref{22-1}), we have the scale factor 
\beq  
a(t)=(\sqrt{2GM_0}\gamma)^{1/\gamma}t^{1/\gamma}.
\eeq
When $\gamma=3/2$, it reduces to the result for the
radiation-dominated flat universe. For large $t$ the comoving distance
to the horizon is 
\beq   
r_H=\int^t_0\dpf{dt'}{a(t')}=\dpf{1}{\sqrt{2GM_0}(\gamma-1)}
(\sqrt{2GM_0}\gamma t)^{(\gamma-1)/\gamma}
\eeq
Therefore, the entropy density has the form
\beq   
\dpf{S}{A}\sim \dpf{\sigma}{\sqrt{2GM_0}(\gamma-1) a^{2-\gamma} (t)}
\eeq
If initially the holographic bound holds, namely
$\sigma /a^2_0 \leq 1$,
then, within the region $1<\gamma\leq 2$, we find again that subsequently the
ratio $S/A$ will always be smaller than unit. This result will also hold
in open universe case $(k=-1)$.

Most interesting  is to see now whether the holographic bound
can be realized in the closed universe $(k=1)$. From (22), we find that
$\dot{a}$ vanishes at $a=(2GM_0)^{1/2(\gamma-1)}$, afterwards $\dot{a}$
becomes negative and the universe collapses. At the turning point, we get the
value of the particle horizon $L_H$
\beq 
L_H(turning)=\dpf{1}{2 (\gamma -1) }
(2GM_0)^{1/2(\gamma -1)}B(1/2,1/2)
\eeq
where $B(p,q)$ is the Euler Beta function. Then we see that at  the
turning point
\beq      
\dpf{S}{A}\sim\dpf{\sigma}{2 (\gamma -1) a(turning)} \label{26-1}
\eeq
Let us impose that  the ratio $S/A$ is smaller than unit
at the turning point. We now  study what happens near the final
stages of collapse when the scale factor reaches the Planck scales. By
symmetry, at this time $L_H\sim \dpf{2a_0}{a(turning)}L_H(turning)\sim
\dpf{a_0}{ \gamma -1}$, and $\sigma/a^2_0 \sim 1$.
Hence the ratio of $S/A$ yields
\beq  
\dpf{S}{A}\sim \dpf{a_0}{ \gamma -1 }
\eeq
In order that the holographic bound  be realized we need
$a_0< \gamma -1 $. However, even if the scale factor at 
the Planck time is small enough the holographic principle 
will not be valid: by the same arguments used before, eq. (\ref{26-1})
says that the bound is violated at the turning point, if
$1 <\gamma \le 2$. 

For other values of $\gamma$ the situations are naively expected to be 
improved. For $\gamma <1$ case in four dimensions, it has been proven 
in \cite{12} that this is the case. However, in three dimensions the 
situation is worse, since for $\gamma <1$ equation (\ref{22-1})
does not accomodate classical solutions immediately after the Big bang.

If we impose $a_0 < \gamma -1 $, we need
\beq \label{29-1}
\sqrt {2GM_0} < \frac {( \gamma -1)^\gamma}{\gamma}
\eeq
and at the turning point, it is easy to see that for $\gamma >2$ we
need
\beq
\frac SA {{>}\atop \sim} \frac {\sigma}2 \frac {\gamma^{-\frac 1{1-\gamma}}}
{\vert \gamma -1\vert^{\frac{2\gamma -1}{\gamma -1}}}
\label{fracsa1}
\eeq
and there is no evident contradiction with the holographic bound for
$\gamma >2$, since the above function decreases from one (at $\gamma =2$)
downwards.

\begin{figure}[ht]
\hspace*{3.5cm}
\scalebox{.5}{\includegraphics*[35mm,143mm][168mm,235mm]{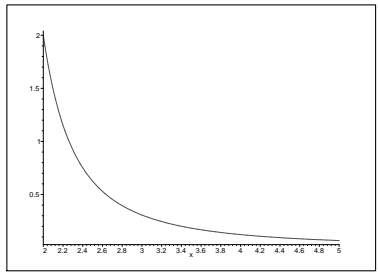}}
\caption[]{\footnotesize Function of $\gamma$ as displayed in the r.h.s of
(\ref{fracsa1}).}
   \label{fig}
\end{figure} 

However, the problem is not that simple. The restriction $a_0 < \gamma -1$ 
together with Eq. (\ref{29-1}) prevents the universe to have enough momentum
to expand sufficiently and the universe would collapse soon after it
was born. Hence the classical description used to discuss the holographic
principle by introducing unusual matter is not valid.

In summary, we have investigated the holographic principle in
(2+1)-dimensional cosmological models. We have shown that in all flat and
open universes, the holographic principle is satisfied in all models. This
result is  consistent with that obtained in (3+1)-dimensional cosmologies. 
For (2+1)-dimensional closed universe the attempts to maintain the
holographic principle  failed, no matter whether the universe is
composed of normal matter, strange negative pressure matter, or strange
matter with high pressure. The result for the closed universe with
negative pressure matter is quite different from that claimed in
(3+1)-dimensional cases [12]. The negative result for holographic
principle in (2+1)-closed universe gives us even more motivation to modify
the holographic principle in cosmology.
 
ACKNOWLEDGEMENT: This work was partically supported by
Fundac$\tilde{a}$o de Amparo $\grave{a}$ Pesquisa do Estado de
S$\tilde{a}$o Paulo (FAPESP) and Conselho Nacional de Desenvolvimento 
Cient$\acute{i}$fico e Tecnol$\acute{o}$gico (CNPQ).  B. Wang would also
like to acknowledge the support given by Shanghai Science and Technology
Commission.

\end{document}